\documentclass[12pt,preprint]{aastex}
\slugcomment{Submitted to Astrophysical J.}
\shorttitle{BH Neutrino}
\shortauthors{Sasaqui et al.}

\newcommand{\ala}{$\alpha$($\alpha$n,$\gamma$)$^{9}$Be}
\newcommand{\anc}{($\alpha$,n)$^{12}$C}
\newcommand{\atg}{$\alpha$(t,$\gamma$)$^{7}$Li}
\newcommand{\lin}{$^{7}$Li(n,$\gamma$)$^{8}$Li($\alpha$,n)$^{11}$B}
\begin{document}
\title{Supernova Neutrino-Effects on R-Process Nucleosynthesis in Black Hole
Formation}
\author{T.\ Sasaqui\altaffilmark{1,2,5}, T.\ Kajino\altaffilmark{1,2} , and
\ A. B. Balantekin\altaffilmark{3,4}}
\altaffiltext{1}{Department of Astronomy, Graduate School of Science,
University of Tokyo, 7-3-1
Hongo, Bunkyo-ku, Tokyo 113-0033, Japan}
\altaffiltext{2}{National Astronomical Observatory and
The Graduate University for Advanced Study,
2-21-1 Osawa, Mitaka, Tokyo 181-8588, Japan}
\altaffiltext{3}{Physics Department, University of Wisconsin
1150 University Avenue
Madison, WI 53706
U.S.A.}
\altaffiltext{4}{Department of Physics, Tohoku University, Sendai
980-8578, Japan}
\altaffiltext{5}{Present address: Naka Division, Nanotechnology 
Products Business Group,
Hitachi High-Technologies Corporation, 882, Ichige, Hitachinaka, Ibaraki
312-0033, Japan}

\begin{abstract}
Very massive stars with
mass $\geq 8 M_{\odot}$ culminate their evolution by supernova 
explosions which are presumed to be the most viable candidates 
for the 
astrophysical sites of  r-process nucleosynthesis.
If the models for the supernova r-process are correct, then
nucleosynthesis results could also pose a significant constraint
on the remnant of supernova explosions, $i.e.$ neutron star or black hole.
In the case of very massive core collapse for progenitor mass 
$20 M_{\odot} \sim 40 M_{\odot}$,
a remnant stellar black hole is thought to be formed.  Intense
neutrino flux from the neutronized core and the neutrinosphere might
suddenly cease during the Kelvin-Helmholtz cooling phase
because of the black hole formation.
It is important and interesting to explore observable consequences of such a
neutrino flux truncation.
Arguments have recently been given in the literature
that even the neutrino mass may be determined from the
time delay of deformed neutrino energy spectrum after the cease
of neutrino ejection (neutrino cutoff effect).

Here, we study the expected theoretical response of the r-process
nucleosynthesis to the neutrino cutoff effect
in order to look for another independent signature of this phenomenon.
We found a sensitive response of the r-process yield
if the neutrino cutoff occurs after the critical time when
the expanding materials in the neutrino-driven wind drop out of
the Nuclear Statistical Equilibrium (NSE).  
The r-process nucleosynthesis yields drastically change 
if the cutoff occurs during the r-process, making maximal effect on the
 abundance change in $^{232}$Th and $^{235,238}$U. There is large
 probability of finding this effect in elemental abundances of r-process
 enhanced metal-deficient halo stars whose chemical composition is
 presumed to be affected by population III supernovae in the early
 Galaxy.  
Using this result, 
connected with future detection of
the time-variation of SN neutrino spectrum, we are able to identify
when the black hole formation occurs in the course of SN collapse.
\end{abstract}


\section{Introduction}

In the past eighteen years after SN1987A emerged in the Large
Magellanic Cloud, a number of interesting scenarios have been proposed
to explain why a pulsar has not been discovered at the supernova
(hereafter, SN)
remnant. One of the most viable scenarios is the possibility
that the pulsation is too weak for emitted X-rays to penetrate through
the ambient gas clouds surrounding the SN remnant. 
Another interesting possibility is that the proto-neutron star
(hereafter, NS)
was first formed after the core-collapse, emitting thermal neutrinos in
twelve seconds during the Kelvin-Helmholtz cooling phase, then became a
black hole (hereafter, BH) after deleptonization \citep{brownbethe94}.

BH formation in core-collapse SNe leads to an 
interesting theoretical study.
If BH formation happens when the flux of neutrinos is still high,
abrupt termination within \(\leq 0.5\) ms would emerge as a sharp cutoff
of neutrino luminosity at some time after the core bounce. This feature
allows celestial model-independent time-of-flight mass tests for the
three light neutrino families \citep{beacom01}.  
\citet{beacom01} discussed 
detectable neutrino
mass sensitivity that depends strongly on the timing, i.e. how early BH 
formation occurs after core collapse.

Motivated by these theoretical studies, we consider the 
possibility that the r-process
nucleosynthesis could provide an independent observable signature
showing BH formation during high neutrino luminosity epoch.
All flavors of neutrinos and antineutrinos from the neutrino-spheres
inside the proto-NS are thought to play at least two
essential roles in successful SN explosions in ``delayed explosion''
model. First, successive interactions between intensive flux of
neutrinos and materials collapsing into the proto-NS deposit
neutrino energy into the ejecta and thus revive the shock
propagation leading to a successful breakout through the iron core
\citep{wilson85,bethewilson85}.  Second, in such a
neutrino-powered SN explosion mechanism, the atmosphere of the 
proto-NS 
is heated by neutrinos at high entropy $s/k$ = $100-400$ to form a
``hot bubble'' flowing out rapidly behind the shock, which is called the
neutrino-driven wind. This is a viable candidate site for the r-process 
\citep{woosley94}.

The evolution of the neutrino-driven wind begins from high temperatures 
about $10^{10}$K at high entropy $s/k$ = $100-400$, thus the system is
in nuclear statistical equilibrium (NSE) and favors free neutrons,
protons, and some amount of $\alpha$-particles. As the temperature drops
below about \(kT \sim 0.5\) MeV, fast charged particle reactions,
which are responsible for interconverting protons into $\alpha$-particles and
converting $\alpha$-particles into composite nuclei, 
very quickly accumulate ``seeds'' 
that have large masses \(70 \leq A \leq 120\) ($\alpha$-process).
This lasts until the temperature drops to \(kT \sim 0.5/e \sim 0.2\) MeV
when the charged particle reactions freeze out ($\alpha$-rich
freezeout). Below this temperature, only neutron-capture flow goes on,
followed by $\beta$-decays, and the r-process occurs until the neutrons
are exhausted (r-process and freezeout).
In Figure \ref{fgcut-temp} we display the evolution of temperature 
(top panel)
and chemical composition of the light elements (bottom panel),
indicating the sequence of nuclear reaction
processes, $i.e.$ the NSE at \(t \leq t_{\alpha}\), the $\alpha$-process
at \(t_{\alpha} \leq t \leq t_{\rm n}\), and the r-process at
\(t_{\rm n} \leq t \leq t_{\rm f}\). Here, we define the beginning of the
$\alpha$-process at \(t = t_{\alpha}\) when \(kT = 0.5\) MeV, the
$\alpha$-rich freezeout at \(t=t_{\rm n}\) when \(kT = 0.5/e \sim
0.2\) MeV, and the r-process freezeout at \(t = t_{\rm f}\).
We also define the dynamical expansion time of the wind, $\tau_{dyn}$,
as the e-fold decay time of the temperature from $kT = 0.5$ MeV \citep{qian96},
i.e. $\tau_{dyn} = t_{\alpha}- t_{\rm n}$.
A short dynamical time can suppress the overproduction of seed nuclei,
leaving plenty of free neutrons for the subsequent neutron-capture flow.

The important neutrino reactions during the nucleosynthesis are
\begin{eqnarray}
\nu_{e}+{}^{Z}_{N}A \rightarrow {}^{Z+1}_{N-1}A+e^{-},\\
\bar{\nu_{e}}+{}^{Z}_{N}A  \rightarrow {}^{Z-1}_{N+1}A+e^{+},\\
\nu_{\rm x}(\bar{\nu_{\rm x}}) + ^{Z}_{N}A \rightarrow \left[
\begin{array}{c}
^{Z-1}_{N}A + \rm{p}\\
^{Z}_{N-1}A + \rm{n}\\
\end{array}
\right]
+ \nu_{\rm x}^{'}(\bar{\nu_{\rm x}^{'}}),
\end{eqnarray}
where x= $\mu$, and $\tau$ are the neutrino flavors, and $^{Z}_{N}A$ 
is the nucleus with proton number Z and neutron number N.
In particular the charged-current reactions that determine the initial
neutron-to-proton ratio are
\begin{eqnarray}
\nu_{e}+\rm{n} \rightarrow \rm{p}+e^{-},\\
\bar{\nu_{e}}+\rm{p} \rightarrow \rm{n}+e^{+}.
\end{eqnarray}
The neutron-to-proton ratio in the weak equilibrium satisfies 
\citep{qian96},
\begin{eqnarray}
Y_{e}={{\rm p}\over{{\rm n}+{\rm p}}} \approx \bigl(
1 + {{L_{\bar{\nu_{e}}}}\over{L_{\nu_{e}}}} \times
{
{\epsilon_{\bar{\nu_{e}}} -2\Delta +1.2 {\Delta}^2/{\epsilon_{\bar{\nu_{e}}}}}
\over{\epsilon_{\nu_{e}} +2\Delta +1.2 {\Delta}^2/{\epsilon_{\nu_{e}}}}
}
\bigr),
\end{eqnarray}
where $Y_e$ is the electron fraction, $L_{\nu}$ is the neutrino 
luminosity of each species,
$\epsilon_{\nu}$ is the average energy proportional to $T_{\nu}$, and
$\Delta$ is the mass difference between proton and neutron.
In this paper we take \(kT_{\nu_{e}}=11.0\) MeV, 
\(kT_{\bar{\nu_{e}}}=19.0\) MeV, and
\(kT_{\nu_{\rm{x}},\bar{\nu_{\rm{x}}}}=25.0\) MeV from \citet{woosley94} and 
\citet{qian97}.
Because of the hierarchy of different neutrino flavors
\(T_{\nu} < T_{\bar{\nu_{e}}} < T_{\nu_{\rm{x}}(\bar{\nu_{\rm{x}}})}\), 
which is
the consequence of the different diffusion length scales and the
decreasing temperature with increasing radius, this $Y_{e}$ value is
less than 0.5 when \(L_{\nu_{e}}= L_{\bar{\nu_{e}}}\), as indicated by the 
numerical simulations of the neutrino transfer \citep{woosley94}.
Therefore, the whole nucleosynthesis sequence described above and
in Figure \ref{fgcut-temp} occurs
in the neutron-rich environment with \(Y_{e} = {{\rm p}\over{{\rm n}+{\rm
p}}} < 0.5\).

Since the timescale of the $\alpha$ production is
shorter than those of the reactions (4) and (5), protons are locked up
into $\alpha$ particles until they are exhausted. Therefore
the reaction (5) does not occur at later times.
Thus, the nuclei are neutron rich and the neutrons in the nuclei 
are degenerate up to higher energy levels than those of the protons.
Hence, the reaction (1), which is fundamentally equivalent to the reaction
(4), is energetically favorable. In contrast, the reaction (2) can not occur 
because of the Pauli exclusion principle.
Both reactions (1) and (4) have an important role in decreasing
the neutron abundance.
If the BH is formed and the neutrino luminosity is cut off,
neutron abundance is kept high. This may lead to an efficient production
of r-process elements because the neutron-to-seed ratio are large under
such high neutron abundance conditions \citep{terasawa01}.
We note that among all neutral-current reactions (3), the
spallation of $\alpha$-particles;
\begin{eqnarray}
\nu_{\rm x}(\bar{\nu_{\rm x}}) + \rm{^{4}He} \rightarrow \left[
\begin{array}{c}
^{3}{\rm H} + \rm{p}\\
^{3}{\rm He} + \rm{n}\\
\end{array}
\right]
+ \nu_{x}^{'}(\bar{\nu_{x}^{'}}),
\end{eqnarray}
is most important to accelerate the termination
of charged particle reactions during the $\alpha$-process by
accumulating more seed nuclei and leaving less free neutrons at the end
of the $\alpha$-process at \(t \approx t_{\rm n}\) \citep{meyer95}. This also
strongly affects the r-process nucleosynthesis.

As such, the neutrino cutoff may affect nucleosynthesis because of robust
emission of neutrinos before the cutoff time $t_{cut}$.
Produced r-process abundance pattern may differ from
that produced in the SNe leading to the formation of NS.
This difference might be important to understand the mechanism of BH 
formation, which remains an open question.

In our calculations we ignore the effects of neutrino-flavor mixing 
\citep{qian95}. Neutrino mixing can significantly alter the r-process 
nucleosynthesis yields (for a recent discussion see \citet{bal05}). 

This paper will be organized in the following way. In sect. \ref{BHFaNC},
we first review various scenarios of BH formation, and then
discuss the conditions under which the neutrino luminosity is cut off in the
manner which we will discuss in the present article. In
sect. \ref{FMNDW}, we describe flow models of neutrino-driven wind in
the core-collapse SNe which we adopt in our numerical studies. Several
assumptions and approximations are explained. We also explain the
nuclear reaction network code in this section. In sect. \ref{RD}, we
show the calculated results on the r-process nucleosynthesis and discuss
how different r-element abundances we would obtain, depending on the
BH vs. NS formation after the core-collapse.
Finally, in sect. \ref{SFO}, we summarize the present paper and
discuss the implications of the spectroscopic
observations of actinide elements, $^{232}$Th and $^{235,238}$U, in
metal-deficient halo stars for testing our results.

\section{Black Hole Formation and Neutrino Cutoff}
\label{BHFaNC}

We calculate the r-process nucleosynthesis in the SNe 
where BH is formed.
Since we do not know when the BH is formed,
we treat the time $t_{cut}$, at which the neutrino is cut
off, as a parameter.
We adopt the constant luminosity, \(L_{\nu} = 10^{51} \rm{erg s}^{-1}\)
at $t \leq t_{cut}$, and assume that
neutrino luminosity is zero after the neutrino cutoff, \(L_{\nu} = 0\) 
at $t_{cut} < t$.
We investigate the dependence of the r-process nucleosynthesis on 
the value of $t_{cut}$.

Stars more massive than 8$M_{\odot}$ are known to culminate
their main sequence lives as core-collapse SNe.
In a standard SN explosion where the  progenitor star has a mass $\sim$
20$M_{\odot}$, in general, a NS is formed as a remnant in the
center. In heavier progenitor stars, a BH can be formed
instead of NS. 
From a cosmological point of view, the recent discovery of an extremely
metal-deficient star suggests that such a
supermassive star that induces a BH existed at least in the
early universe \citep{frebel05,iwamoto05}.

There are so far two different scenarios for BH formation 
in SN explosions.
One of them is a collapsar model. This model is based on a completely
different picture from the standard SN. In this model a BH is
thought to be formed immediately after a core collapse of SNe.
It is phenomenologically possible when the mass of the
progenitor star is heavy enough ($e.g. \geq$ 40$M_{\odot}$) to produce
a BH.
An accretion disk is formed around a BH by pulling materials into
the central region and strong X-rays
or $\gamma$-rays are emitted. This scenario is sometimes invoked 
to explain $\gamma$-ray bursts \citep{mac99}.

A second scenario is based on a standard core collapse SN explosion.
If the proto-NS mass exceeds the maximum NS mass,
then a BH can be formed during the SN explosion.
Here the maximum mass is thought to be 2.2$M_{\odot}$
\citep{beacom01,akmal98}.
This scenario can be useful for the progenitor mass region heavier
than that of a standard SN and lighter than that of a collapsar
($e.g.$ between 20$M_{\odot} \sim$ 40$M_{\odot}$).
It is the scenario for which the neutrino cutoff effect by the BH 
formation is obvious, since there can exist a proto-NS $i.e.$
neutrino flux from the central neutrino sphere for a short time until
a BH is formed.
This situation is different from the collapsar model.

\section{Flow Models of Neutrino-Driven Wind}
\label{FMNDW}

\subsection{Background}

The flow dynamics in non-spherical SN explosion is complicated
in the case of collapsars \citep{mac99} or hypernovae \citep{maeda03}
which are associated with BH formation.
The flow in which BH is formed might be different from the
standard SN flow leading to a NS formation.
We however assume similar models for both flows for the following reasons.
The first reason is that in this work we are mainly interested in the 
consequence of the neutrino-cutoff, i.e. how it destroys neutron richness and
affects the final abundance of r-process elements.
Since the r-process condition in Type-II SN explosion leaving NS as a remnant
has been studied very well \citep{woosley94,witti94,otsuki00},
it is effective to extensively study the r-process in BH formation
in similar models by tuning flow parameters of  the neutrino-driven wind.

The second reason is that the luminosity stays at
high value of $10^{51} \sim 10^{52}$ erg$\rm{s^{-1}}$ for each neutrino species
before the neutrino-cutoff at 1-2 s in either models of BH formation
as discussed in the previous section.
As we will discuss below, neutrino heating energy is so efficiently deposited
in very short period $\sim 3$ ms \citep{otsuki00} that the hot bubble
may form even in the BH formation.

The third reason is that we have not yet obtained a realistic 
theoretical simulation
of a successful SN explosion except for the model of  \cite{wilson85}.
It may be more difficult to simulate BH formation
which needs at least two dimensional numerical analysis including the effects
of general relativity in order to describe both the dynamics of the 
accretion disk and jet formation as well as the inner core collapse into BH.

For these reasons we use the approximation of the neutrino-driven
wind which is suitable for our purpose to study the neutrino-cutoff effects
on the r-process nucleosynthesis.

\subsection{Steady-State Flow Model}

We adopt the spherical steady-state 
flow model for the neutrino-driven wind
\citep{qian96,takahashijanka97,otsuki00,wanajo01}.
This flow model is one which leads to a successful
r-process.
Even though the entropy per baryon is moderately low,
$s/k\approx$ 100-300, the r-process can occur in this
neutrino-driven wind when the dynamical expansion timescale becomes much
shorter than the collision timescale of neutrino-nucleus interactions.
For the present application, such hydrodynamic flow can be
approximated \citep{otsuki03} by solving the following
non-relativistic equations:
\begin{eqnarray}
4 \pi r^2 \rho v = \dot{M},\\
{1\over 2}v^2 - {{GM}\over{r}} + N_{A}~s_{rad}~ \mathrm{kT}= E,\\
s_{rad} ={{11 {\pi}^2}\over{45\rho N_{A}}}\left({{\mathrm{k}T}\over{\hbar
c}}\right)^3,
\end{eqnarray}
where $\dot{M}$ is the rate at which matter is ejected by neutrino
heating on the surface of the proto-NS.
In Eq. (9), the total energy $E$ is
fixed by the boundary condition on the asymptotic temperature,
$\mathrm{T}_{a}$;
\begin{eqnarray}
  E=N_{A}~s_{rad}~\mathrm{k}~\mathrm{T}_{a},
\label{energy}
\end{eqnarray}
and we take into account only photons and $e^{\pm}$ pairs
in the estimate of the entropy per baryon, $s_{\rm{rad}}$, in Eq. (10).
For simplicity, in the present work, we utilize an adiabatic,
constant-entropy wind rather than computing neutrino heating
explicitly \citep{otsuki00}.
However, we include both charged- and neutral-current interactions
between neutrinos and nuclei in the nucleosynthesis. A constant
neutrino luminosity $L_{\nu}=10^{51} \> \rm{ergs^{-1}}$ for each neutrino
species is also adopted.
This model has four parameters, which are the NS 
mass, entropy, boundary temperature, and mass loss rate.

For the purpose of investigating many different flow models,
we use the exponential model \citep{otsuki03} which satisfies 
\begin{eqnarray}
\tau_{dyn}^{-1}=-{{1}\over{T-T_{a}}}{{{\rm d}T}\over{{\rm d}t}} 
\label{tau_dyn_def},
\end{eqnarray}
where $T_a$ is the asymptotic temperature as defined in Eq. (11)
so that the solution can be a good approximation to the exact solution of
the set of Eqs. (8) - (11).
Under the approximate condition that the entropy
  (\(\mathrm{s}=({{11 {\pi}^2}/{45 \mathrm{N_{A}}}})(
{{T^{3}}/{\rho}})\)) is constant,
temperature and density are given for a fixed $\tau_{dyn}$.
Among many different flow models thus solved, we here
adopt typical flows with short dynamical explosion time scales,
$\tau_{dyn} \leq 50 \> \rm{ms}$, by varying the entropy:
  ($\tau_{dyn}$ [msec], $s/k$)=(5, 355), (10, 480), (20, 750),
(30, 1050), and (50, 1680), where we fixed the other parameters.
These models produce almost the same final abundance
of r-process elements \citep{sasaqui05c} and can well explain
the abundance pattern of neutron-capture elements
which were detected in one of the most r-process element enhanced
metal-deficient halo star CS22892-052 \citep{sneden96}.

\subsection{Neutrino Heating and Cutoff Time}

\citet{otsuki00} showed that neutrino heating occurs most effectively at
$r \leq 30$ km from the center of the collapsing core.
We found in the same flow model analysis that
it takes $3-4$ ms for the material blowing off the surface from 
proto-NS to reach 30 km.
We discussed in sect. 2 that
neutrino emission is followed until abruptly terminated by the BH formation
with the appearance of apparent horizon in the first scenario.
Finite time elapses before BH forms, and the neutrino luminosity is 
cutoff at $1-2$ s \citep{burrows88}.
In the second scenario the cutoff time is
delayed about 10 s for the formation of proto-NS, followed by
possibly a phase transition of softening the EOS of core matter, and
possibly later mass accretion onto BH.
Since the neutrino cutoff time $1-2$ s or 10 s is larger than
the heating time scale $3-4$ ms,
we can assume that the hot bubble may form in the neutrino-driven wind
of Type-II SNe that form BH as a remnant.

The temperature of all flows of the neutrino-driven wind used in the 
present study 
reaches to $T_9 \approx 9$ at $t = 3-4$ ms.  We adopt this temperature
as the typical temperature at which neutrino heating is completed. The
entropy of the hot bubble formed stays constant after this time.
This justifies our assumption of constant entropy during the 
nucleosynthesis which follows.
Our nucleosynthesis calculation starts from this initial temperature,
and time zero refers to the time when the hot bubble reaches $T_9 \approx 9$
as displayed in Figure \ref{fgcut-temp}.
It is to be noted that the time in this figure and all others we discuss later
is not the time after core collapse or core bounce.
Flows of the neutrino-driven wind successively blow off until the 
neutrino luminosity
is cutoff at the time $t = t_{cut}$.

\section{Nucleosynthesis Network}

For the calculation of the r-process nucleosynthesis, we employ the reaction
network used in \citet{sasaqui05a,sasaqui05b}, which was developed from the
original dynamical network code calculations described in
\citet{meyer92},
\citet{woosley94}, \citet{terasawa01}, and \citet{otsuki03}.

The main feature of the network code developed 
\citep{sasaqui05a,sasaqui05b} is the
improvement in the light-mass nuclear reactions. It has already been
discussed in literature \citep{woosleyhoffman92,meyer92,woosley94} 
that the \ala \anc~
reaction sequence is particularly important in the earlier stage of the
$\alpha$-process at high temperature and high density for the production
of r-process seed nuclei. \citet{terasawa01} suggested that as
long as the expansion timescales of the neutrino-driven winds are 
short, \(\tau_{dyn} \le 10\)ms, a successful r-process occurs 
\citep{otsuki00}. Another nuclear reaction-flow path along the
neutron-rich unstable nuclei may play as significant a role as the \ala~
reaction does. \citet{sasaqui05a,sasaqui05b} found that this is always the case
regardless of the flow models of neutrino-powered SN explosions. They
quantitatively identified that the \atg~  reaction and the subsequent
\lin~ reaction are the most critical reactions in addition to \ala  \anc~
that affect strongly the r-process nucleosynthesis particularly of
the actinide elements, $^{232}$Th and $^{235,238}$U. 
They also found that these different nuclear reaction paths merge at
$^{14}$C, which is followed by neutron-capture flows on carbon, nitrogen
and oxygen isotopes to manifest a new feature of ``semi-waiting'' point
at the neutron-rich isotopes $^{16}$C, $^{18}$C, and $^{24}$O. This
feature of the ``semi-waiting'' point is the manifestation of a primary
r-process so that the SN nucleosynthesis starts from the high entropy
conditions on which the NSE favors neutrons, protons, and some amount of
$\alpha$-particles as an initial composition and thereby the
$\alpha$-capture, neutron-capture, and $\beta$-decay compete with one
another in the light-mass neutron-rich nuclei.  \citet{sasaqui05a,sasaqui05b}
updated many nuclear reaction rates on light-mass neutron-rich nuclei
with the help of recently accumulated new experimental data obtained by
using radioactive nuclear beams \citep{nakamura99,sasaqui05a,sasaqui05b}
and references therein.

We also note that we calculate the nucleosynthesis sequence from the
NSE, $\alpha$-process, $\alpha$-rich freeze-out, r-process, and subsequent
beta-decay and alpha-decay, as explained in sect. 1 and in Figure
\ref{fgcut1},  in a single network code rather
than to split the calculation into two parts as was done in
\citet{woosley94}. This is important for our present 
study in looking the observable signatures of BH formation in 
elemental abundances from the r-process nucleosynthesis.

\section{Result and Discussions}
\label{RD}

R-process nucleosynthesis in neutrino-driven wind of Type-II SN explosions
has been studied by several authors 
\citep{woosley94,witti94,meyer95,qian96,otsuki00}.
The following two conditions prove to be important for a successful r-process:
\begin{eqnarray}
\tau_{dyn} \ll \tau_{\nu},\\
  \tau_{dyn} \ll \tau_{\alpha \alpha n},
\end{eqnarray}
where \(\tau_{\nu} \approx 0.201 \times \rm{L}_{\nu,51}^{-1}
({{\epsilon_{\nu}}\over{\rm{MeV}}})({{r}\over{100\rm{km}}})^{2}({{\langle
\sigma_{\nu}\rangle}\over{10^{-41} \rm{cm}^2}})^{-1}\)s
is the neutrino collision time scale \citep{qian97},
\(\tau_{\alpha \alpha n} \approx [\rho ^2
Y_{\alpha}^{2} Y_{n}
\langle \sigma
v_{\alpha \alpha n}\rangle N_{A}^2]^{-1} \)
is the typical nuclear reaction time scale for $^4$He$(\alpha n,\gamma)^9$Be
which is the slowest one among all charged particle reaction paths,
$\rm{L}_{\nu,51}$ is the
neutrino flux normalized in units of $10^{51}$ erg$\rm{s^{-1}}$, $\epsilon_{\nu}$ is the
averaged electron-type neutrino energy $\sim$ 11 MeV, $r$ is the distance of
r-process site from the center of the core, $\langle \sigma_{\nu} 
\rangle$ is the
averaged cross section
over the neutrino energy spectrum, $\rho$ is the density of the mass
element, and $N_{A}$ is Avogadro's number.
The former relation (13) describes the condition in which the neutron
abundance is kept high when the neutrino process becomes ineffective
(see discussion in the introduction),
and the latter relation (14) describes the condition in which the
neutron-to-seed ratio ($Y_{n}/Y_{seed}$) is high enough to realize
a successful r-process of heavy elements from the seed nuclei.
$Y_{seed}$ is defined as the total seed abundance,
i.e. $Y_{seed} = \sum Y_A$ for 70 $\leq$ A $\leq$ 120.

\begin{figure}
\centering
  \rotatebox{0}{\includegraphics[width=13.5cm,height=9.5cm]{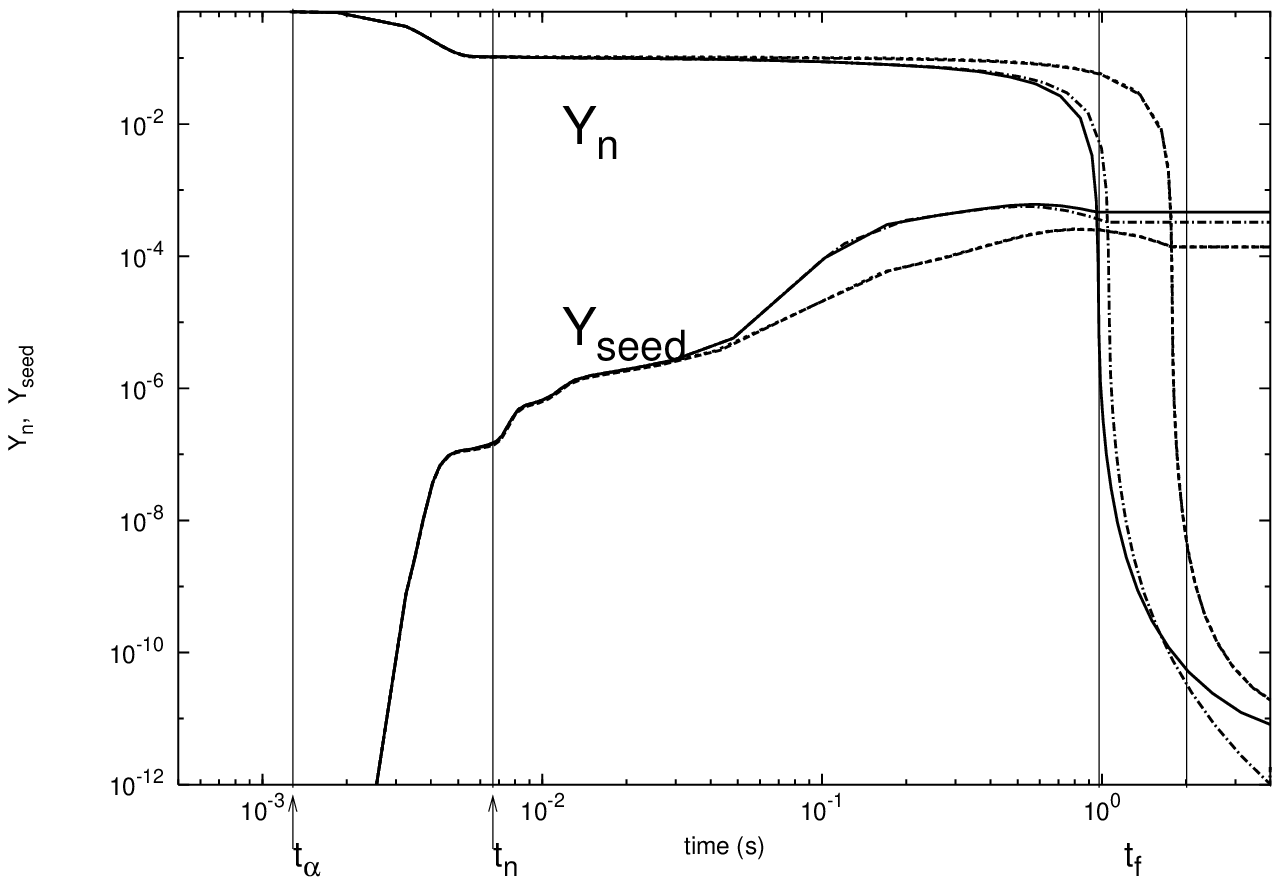}}
  \rotatebox{0}{\includegraphics[width=13.5cm,height=9.5cm]{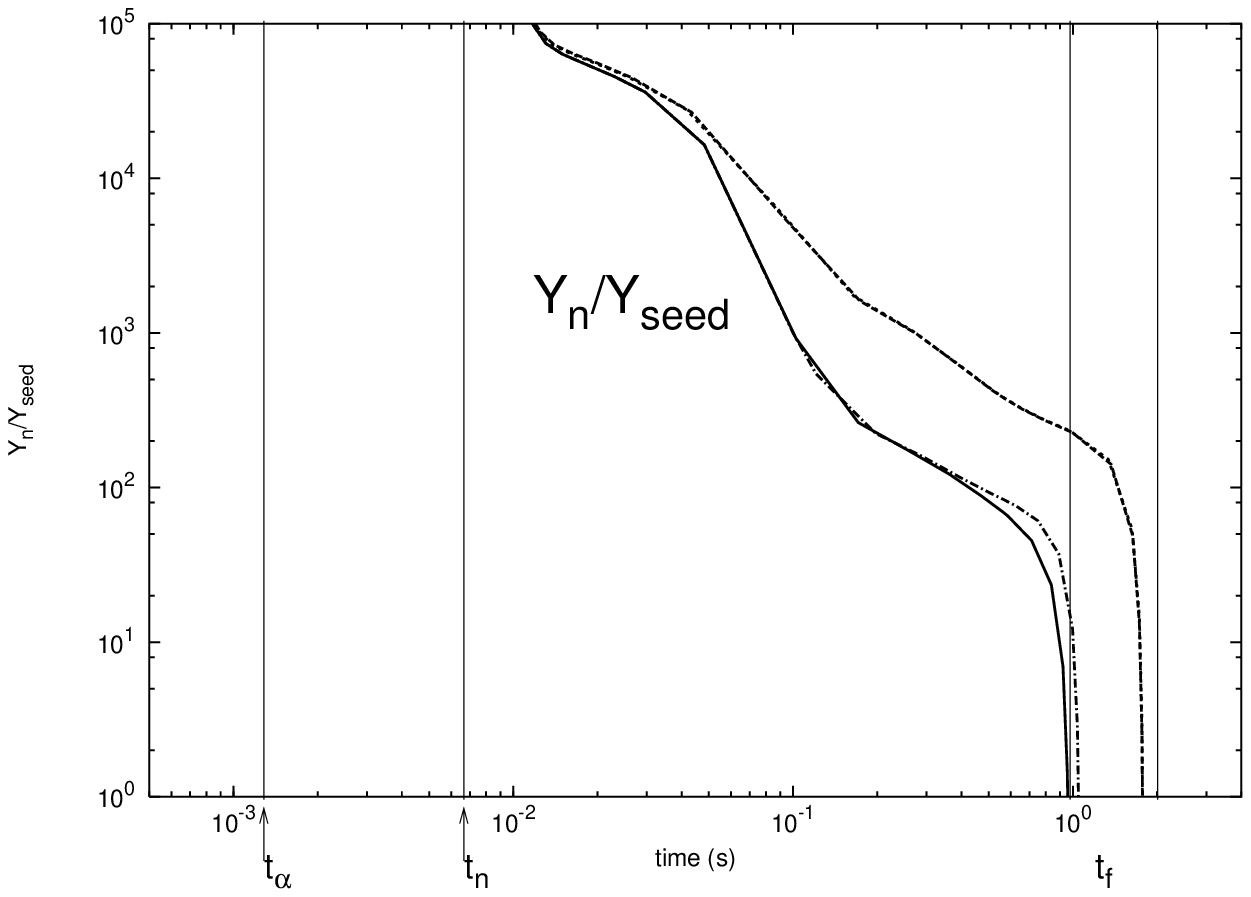}}
\caption{Time evolution of neutron abundance $Y_{n}$ and seed abundance
Y$_{seed}$ (upper panel), and neutron-to-seed ratio
  $Y_{n}/Y_{seed}$  (lower panel) in Otsuki flow model \citep{otsuki00}.
The cutoff times are set to be 0.001s (dashed line),
0.005 s (dotted line),
0.1 s (dash-dotted line),
and $\infty$ (solid line), the same as those in Figure 1.}
\label{fgcut-y}
\end{figure}

The time evolution of the temperature $T_9$, neutron separation energy $S_n$,
and neutron and seed abundances $Y_{\rm{n}}$, $Y_{\rm{seed}}$, and
their ratio $Y_{n}/Y_{seed}$ is shown in Figure \ref{fgcut-temp} and 
Figure \ref{fgcut-y},
separately.
Here, $S_{n}$ means optimum neutron separation energy.
This value is convenient to identify when the classical neutron-capture flow
proceeds at low and almost constant value, $S_{n} \approx 1$ MeV,
and when the freezeout of the n-capture process occurs.
Assuming that the transition probabilities for (n, $\gamma$) and ($\gamma$,n)
are equal to each other and nuclear reaction flow stays at a certain nucleus,
$S_{n}$ is given by
\begin{eqnarray}
S_{n}={{T_{9}}\over{5.040}}\{34.075-\log{(Y_{n}\rho N_{A})}+{3\over 
2}\log{T_{9}}\}.
\end{eqnarray}
When $Y_{n}$ is large in neutron-rich environment, $S_{n}$ is generally
small. A drastic increase in $S_{n}$ gives us the information on neutron
consumption at $\sim$ 1 s as shown in the bottom panel of 
Figure \ref{fgcut-temp}.

\begin{figure}
\centering
\rotatebox{0}{\includegraphics[width=15cm,height=12cm]{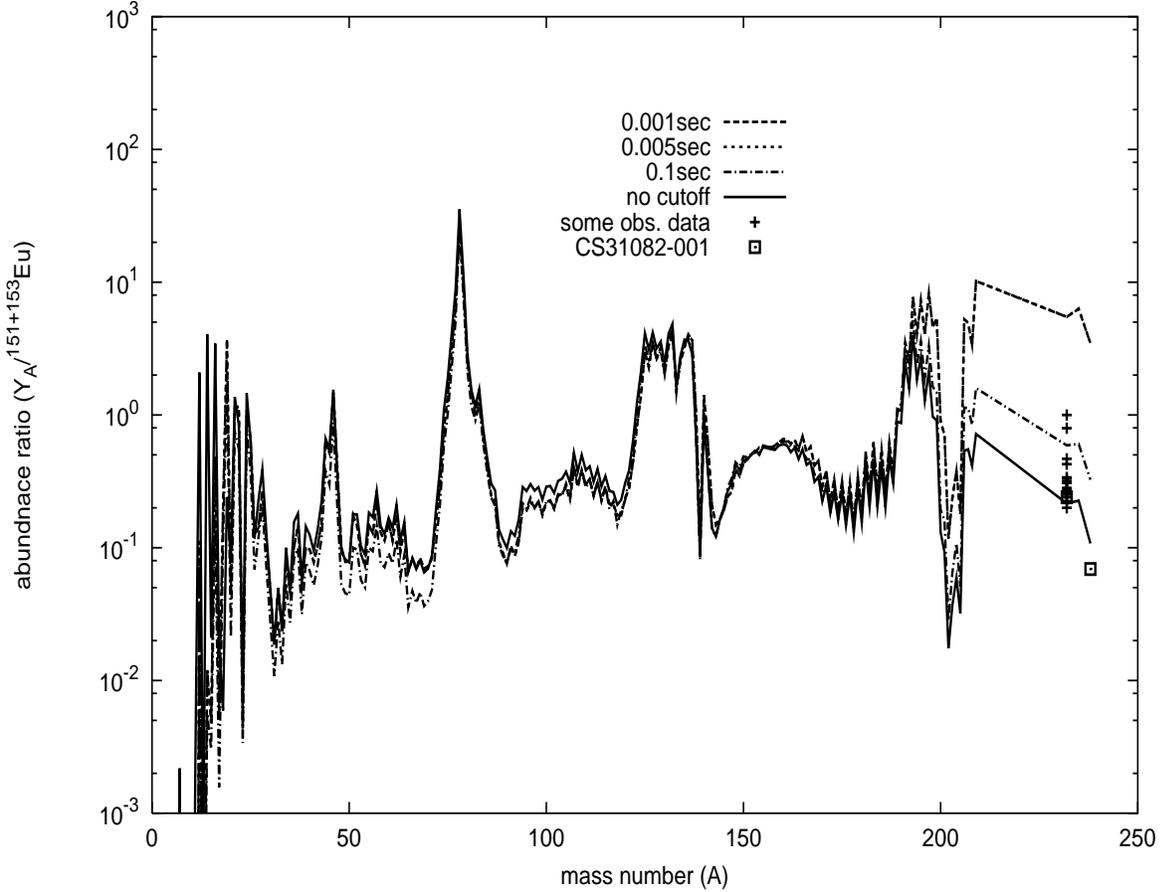}}
\caption{Final abundance patterns as a function of mass number
in the four cases of different neutrino cutoff times calculated in
 Otsuki flow  
model \citep{otsuki00}.
The cutoff times are set to be 0.001s (dashed line),
0.005 s (dotted line),
0.1 s (dash-dotted line),
and $\infty$ (solid line), the same as those for $s_{n}$ in Figure 1.
The points indicate recent observational data of
abundance ratios  $^{232}$Th/($^{151}$Eu+$^{153}$Eu)
and  ($^{235}$U+$^{238}$U)/($^{151}$Eu+$^{153}$Eu) of metal-deficient
 halo stars tabulated in Table 3. }
\label{fgcut1}
\end{figure}

The calculated final abundance patterns are shown
in Figure \ref{fgcut1} for various values of neutrino cutoff time
$t_{cut} = 0.001, 0.005, 0.1$ s, and $\infty$.  The last case is for 
no neutrino cutoff.
The result is summarized as follows:

\noindent
(1) When the neutrino cutoff occurs at $t_{cut} = 0.001$ s
under NSE condition (see Figure \ref{fgcut-temp}),
the influence of the cutoff is largest
as neutron abundance $Y_n$ is kept high (Figure \ref{fgcut-y})
and the r-process becomes effective.
This is because the neutrino interactions on neutrons
occur for a short time before $t_{cut}$.
As shown by the dashed line in Figure \ref{fgcut1}, the heavy 
r-process elements such as
actinides are very effectively produced.
However, in this case the cutoff time $t_{cut} = 1$ ms is too short
for shock wave to break out the iron core and SN might fail to explode
due to the lack of neutrino heating.

\noindent
(2) When the neutrino cutoff occurs at $t_{cut} = 0.005$ s
under the condition of efficient $\alpha$-process (see Figure 
\ref{fgcut-temp}),
the effect of the neutrino cutoff is still large.
The dotted line in Figure \ref{fgcut1} shows that the r-process proceeds
as effectively as in the first case, and the dotted and dashed lines 
are too close
to each other to be separately read out.
Both neutron abundance $Y_n$ and the neutron-to-seed ratio $Y_{n}/Y_{seed}$
in Figure \ref{fgcut-y} are almost the same as those in the first case.

\noindent
  (3) When the neutrino cutoff occurs at $t_{cut} = 0.1$ s
under the condition of efficient neutron-capture process (see Figure 
\ref{fgcut-temp}),
the effect of the neutrino cutoff is small,
and the final r-process abundance denoted by the dash-dotted line in
Figure \ref{fgcut1} is very close to the result of no neutrino cutoff 
(solid line).
A notable difference from the previous two cases (1) and (2) is seen
in $Y_n$ and $Y_{n}/Y_{seed}$ of Figure \ref{fgcut-y}
for $t_{\rm{n}} < t_{cut}$.

\noindent
(4) When we do not take account of the neutrino cutoff for $t_{cut} = \infty$,
calculated $^{232}$Th abundance is close to the observed lower limit
from several metal-deficient halo stars as summarized in Table 3.
Both $Y_n$ and neutron-to-seed ratio $Y_{n}/Y_{seed}$
in Figure \ref{fgcut-y} are close to those of the case (3).

The neutrino cutoff occurring in various nucleosynthesis stages shown
in Figure \ref{fgcut1} can thus make a remarkable effect
on the actinide production.
This is an important key feature to identify which SN remnant, BH or SN, 
forms in the gravitational core-collapse Type-II SN explosions.

\begin{figure}
\centering
  \rotatebox{0}{\includegraphics[width=15cm,height=12cm]{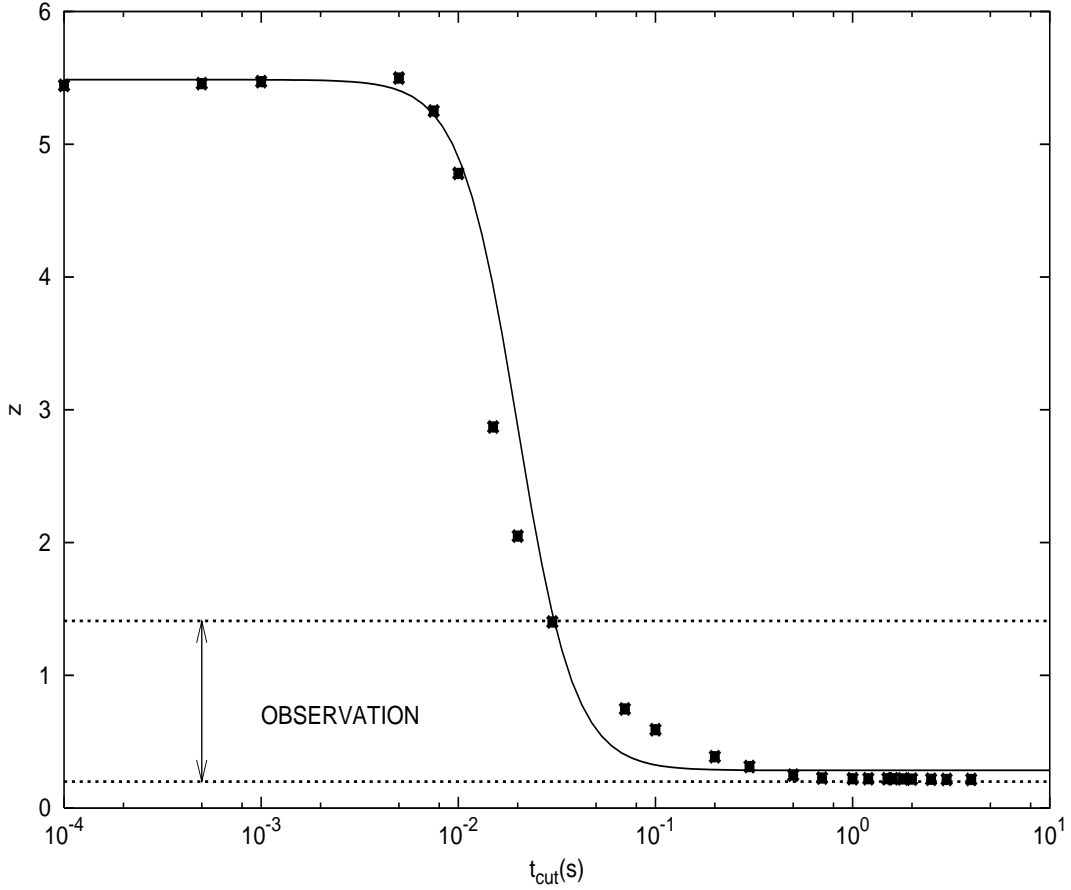}}
\caption{Abundance ratio $z \equiv ^{232}$Th/($^{151}$Eu+$^{153}$Eu)
vs. neutrino cutoff time $t_{cut}$ calculated in Otsuki flow model
 \citep{otsuki00}. Calculated data points shown in crosses are fit by
 the solid curve of Eq.(16) with $\alpha$ = 3.00.
The BH outcome (for $t_{cut} \leq $0.01) and the NS outcome ($t_{cut} 
\geq $0.1)
appear in two different limits of $t_{cut}$.
Drastic change occurs when we set the neutrino cutoff time
right after the $\alpha$-process freezes out and the n-capture process starts.
Observed range include both observational errors and dispersion from
 Table \ref{table4}. 
See Table 1 and Figure 5 for more details.}
\label{fgcut-ra}
\end{figure}

We pay special attention to the behavior of the actinides
($^{232}$Th, $^{235,238}$U) because of their importance in the
cosmochronology. Table \ref{table2} and Figure \ref{fgcut-ra} show
the ratio of $^{232}$Th/($^{151}$Eu+$^{153}$Eu).
This is a useful quantity because astronomical observations
of metal-deficient halo stars cannot provide each isotopic abundance
but just this ratio.
We find in Figure \ref{fgcut1} that the early cutoff time makes this
ratio large. We also find that there appear two
outcomes separated by a narrow transitional neutrino cutoff  
time between 0.01 s
and 0.1 s.  The ``BH outcome'' shows a very high abundance ratio which
is thought to arise from the events of BH formation,
while the ``NS outcome''  shows a low abundance ratio which may
arise from the conditions that do not lead to BH formation.
Also, as discussed in sect. 1, the effect of the
neutrino cutoff by BH formation is mainly due to the change
of neutrons into protons by the weak reaction process (4).
So, the drastic change of the final abundance pattern
occurs if  the neutrino cutoff occurs at a time right after
the end of the $\alpha$-process  and the beginning of the neutron-capture
process, i.e. $t_n \lesssim t$.
This is because drastic environmental change occurs after the
$\alpha$-rich freezeout of making seed elements at $t_n \lesssim t$
and only abundant neutrons are easily affected by the weak process (4).

This profile can be fit by the following function.
\begin{eqnarray}
  z = z_{t \rightarrow \infty} +
{{z_{1}}\over{1+({{t}\over{t_{0}}})^{\alpha}}},
\label{fit1}
\end{eqnarray}
where $z \equiv ^{232}$Th/($^{151}$Eu+$^{153}$Eu),
$t \equiv t_{cut}$, $z_{t \rightarrow \infty}$ means the abundance ratio of no
neutrino cutoff, $z_{1} + z_{t \rightarrow \infty} (\equiv z_{0})$ means the
abundance ratio with no neutrinos,
$t_{0}$ means the start of the drastic change of the ratio $z$,
and $\alpha$ is a constant value. The quantity $\alpha$ is expected to be model
independent and the other quantities are model dependent.
In case of Figure \ref{fgcut-ra}, for example, $z_{t \rightarrow
\infty}$ is 0.284, $z_{1}$ is 5.20, $t_{0}$ is 0.0199 s, and $\alpha$ is
3.00 (Figure \ref{fgcut-ra}) in Otsuki model \citep{otsuki00}.

\begin{figure}
\centering
  \rotatebox{0}{\includegraphics[width=15cm,height=12cm]{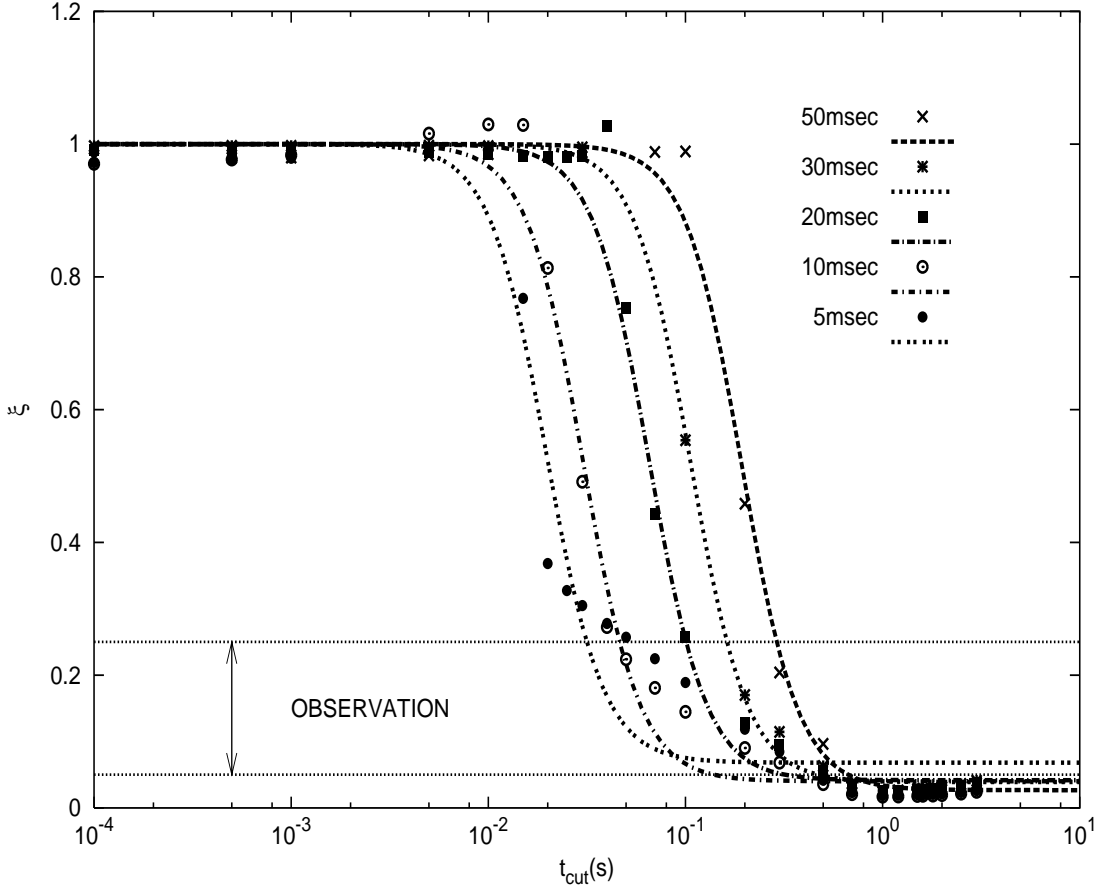}}
\caption{Normalized abundance ratio $\xi = {{z-z_{t \rightarrow 
\infty}}\over{z_{1}}} = {{1}\over{1+(t/t_{0})^{\alpha}}}$
vs. neutrino cutoff time $t_{cut}$ calculated in the exponential models which are
characterized by the different dynamical expansion time scales
$\tau_{dyn} = 5, 10, 20, 30$, and 50 ms of the neutrino-driven wind, 
as indicated.
Calculated data points are fit by Eq. (17) with various parameter values in Table 2.
The parameter $t_{0}$, which corresponds to the time around which
a drastic change of  the final abundance occurs, is
proportional to the dynamical timescale $\tau_{dyn}$.
Observed range includes both observational errors and dispersion from
 Table \ref{table4}. 
See Table 2 and the text for more details.}
\label{fgcut-ratot}
\end{figure}

We repeated this calculation using the exponential flow models which are
characterized by the different dynamical expansion time scales
$\tau_{dyn} = 5, 10, 20, 30$, and 50 ms, as defined in sect. 3.2.
The calculated results are shown in Table \ref{table3}
and Figure \ref{fgcut-ratot} in normalized form of Eq. (16)
\begin{eqnarray}
\xi = {{z-z_{t \rightarrow \infty}}\over{z_{1}}} = 
{{1}\over{1+(t/t_{0})^{\alpha}}}.
\label{fit2}
\end{eqnarray}
Here, we set $\alpha$, the
parameter which determines the shape of fitting function, to be 3, 
the same as
that of Figure \ref{fgcut-ra}.
The parameter $t_{0}$ in Eq. (16), which corresponds to the time around which
a drastic change of  the final abundance occurs, is
proportional to the dynamical timescale $\tau_{dyn}$.
In the models with a small dynamical timescale,
the temperature drops rapidly and
the charged particle reactions are suppressed soon.
The freeze out of the $\alpha$-process occurs early, so the neutrino 
cutoff effect leading to 
a drastic change in the abundance ratio $z$ occurs early.
On the other hand, in the models with a large
dynamical timescale, the temperature drops slowly and the charged
particle reactions continue to take place. The time of the freezeout of the
$\alpha$-process becomes later than that in the models with the small 
dynamical timescale.
The drastic change occurs mostly for this reason.

\citet{beacom01} proposed that the direct possible signature of BH
formation could be the observation of sharp cutoff in the neutrino
signal. This may not be a unique good method because there is increasing
evidence that the BH formation must be rare at the present epoch for
ordinary core-collapse SN scenario, given the low frequency of nearby
SNe. \citet{strigari05} have shown that direct measurements of the
evolution of core-collapse SN rate, which includes the SN rate for BH
formation, are consistent with the predictions based on a variety of
star formation indicators. From the view point of Galactic chemical
evolution, however, different physics may operate \citep{heger03} in the
SN explosions in various environments with very different metalicities
from the early Galaxy to the present epoch. Neither the neutrino
detection studies nor the star formation rate studies can address the
possibility that BH formation was a more frequent and dynamically
important process in the very early Galaxy of active star formation
epoch. 

The nucleosynthesis signal proposed in this paper reveals the promising 
possibility that the
effect of neutrino cutoff associated with BH formation would
manifest itself in the fossil record of the produced r-process
yields. 
Following remarkable advances in spectroscopic observations, 
significant information about the abundance of the r-process elements  
in metal-deficient halo stars
has recently been accumulated. 
These stars are presumed to be second
generation stars which were born in the early Galaxy and their elemental
abundances were ideally affected by a few SN episodes of the
first generation massive population III stars. 

In Figure \ref{fgcut1} we show the observed abundance ratios
$^{232}$Th/($^{151}$Eu+$^{153}$Eu) and\\ 
($^{235}$U+$^{238}$U)/($^{151}$Eu+$^{153}$Eu) of several metal-deficient
halo stars tabulated in Table \ref{table4}. It is interesting that the
typical predicted abundances are near the low end of the observed yield
in this Figure. This might possibly indicate that those actinides were
produced in the first generation SNe associated with remnant NS
formation or indicate that only a partial contribution from the SN
products associated with BH formation is admixed in the observed
r-process elements. The same result is more clearly shown in Figure
\ref{fgcut-ra} and \ref{fgcut-ratot} which display a band of observed
abundance ratios near the ``NS outcome'' rather than the ``BH
outcome''.

\section{Summary and Future Outlook}
\label{SFO}

We investigated the r-process nucleosynthesis in the SNe where
BH could form.
We found that the r-process abundances could change significantly 
by neutrino
cutoff at the BH formation.
This will be one of the predictions about BH formation
if metal-deficient halo dwarfs which have such a specific abundance 
pattern can be found.
There appear no observational signatures at the moment indicating 
that the abundance pattern is made in the events
where the BH can be formed.
Future observations of more halo stars exhibiting enhanced r-process elements
are highly desirable.

We assumed steady state flow of the neutrino-driven wind in our present study.
Actually, the flow model must be effected by the BH formation.
Therefore reliable models with realistic numerical simulations
of general relativistic hydrodynamics are needed. 
However, the dynamics would be more complex since
a very massive star, which 
evolves to a standard SN,
should collapse by accreting materials from the accretion disk and 
eject part of them
in a jet-like explosion.
We need to know  more details about the dynamics where the BH can be formed.
We also need to understand the behavior of the neutrino-driven wind
in the environments where the neutrino luminosity is cutoff.

Massive stars associated with BH formation
are predicted to occur in the early stage of the Galactic evolution so
that 
they could have ejected nucleosynthesis products which are very different from
those ejected from ordinary SNe that leave NS as a remnant. 
In this article we proposed that actinides could show a remarkable
 difference for the different neutrino cutoff effects in SNe leaving BH
 or NS. Similar differences would be found also in the elemental
 abundances of $^7$Li and $^{11}$B \citep{yoshida04,yoshida05} which are
 produced in the neutrino processes in the outer layers of core-collapse
 SNe. 
Such productions could impact the Galactic chemical evolution,
a point worth investigating.
The ultra metal-deficient stars which have such an abundance pattern
might be found in the future using more sophisticated observational
techniques. 

\clearpage


\begin{table}
\centering
\caption{The abundance ratio, $^{232}$Th/($^{151}$Eu+$^{153}$Eu),
vs. the neutrino cutoff time, $t_{cut}$ in units of s, in the flow model
of the neutrino-driven wind \citep{otsuki00} which is characterized by the
dynamical expansion timescale $\tau_{dyn} = 5$ ms.}
\makebox[6.5cm]{\def\arraystretch{0.4}
\begin{tabular}[t]{cc}\hline
{cutoff
  time}&{$^{232}$Th/($^{151+153}$Eu)}
\\\hline \hline
{0.0001}&{5.44}
\\ 
{0.0005}&{5.46}
\\ 
{0.0010}&{5.46}
\\ 
{0.0050}&{5.50}
\\ 
{0.0300}&{1.40}
\\ 
{0.0700}&{0.75}
\\ 
{0.1000}&{0.59}
\\ 
{0.2000}&{0.39}
\\ 
{0.3000}&{0.31}
\\ 
{0.5000}&{0.25}
\\ 
{0.7000}&{0.23}
\\ 
{1.0000}&{0.22}
\\ 
{2.0000}&{0.22}
\\ 
{4.0000}&{0.22}
\\ 
{no cutoff}&{0.22}
\\ \hline
\end{tabular}
}
\label{table2}
\end{table}

\begin{table}
\centering
\caption{The calculated parameter values, $z_{t \rightarrow \infty}$,
$z_{0} (=z_{t \rightarrow \infty} + z_{1})$, $t_{0}$, and $\alpha$,
for the fit of the abundance ratio
$z= ^{232}$Th/($^{151}$Eu+$^{153}$Eu) in the functional form
of Eq. (16) defined in the text.
Five different exponential flow models of the neutrino-driven wind,
which are characterized by the
dynamical expansion timescales $\tau_{dyn} = 5, 10, 20, 30$, and 50 ms,
are adopted.}
\makebox[6.5cm]{\def\arraystretch{0.4}
\begin{tabular}[t]{ccccc}\hline
{$\tau_{dyn} msec$}&{$z_{t \rightarrow 
\infty}$}&{$z_{0}$}&{$t_{0}$}&{$\alpha$}\\\hline
\multicolumn{5}{c}{$^{232}$Th/($^{151}$Eu+$^{153}$Eu)}\\ \hline
{50}&{0.152}&{5.68}&{0.140}&{3.00}\\ 
{30}&{0.231}&{5.90}&{0.106}&{3.00}\\ 
{20}&{0.269}&{6.43}&{0.066}&{3.00}\\ 
{10}&{0.374}&{9.31}&{0.030}&{3.00}\\ 
{5}&{0.483}&{9.00}&{0.023}&{3.00}\\ \hline

\end{tabular}
}
\label{table3}
\end{table}

\clearpage
\begin{table}
\centering
\caption{The list of observational data of actinide elemental
 abundances. $\log_{\epsilon}$(A) is equal to 
  $\log$[N(A)/N(H)] +12, where N(A) is the number abundance of A and N(H)
  is the number abundance of hydrogen. Eu
  is the total abundances of $^{151}$Eu and $^{153}$Eu, and U is the
 total abundances of $^{235}$U and $^{238}$U. Observational errors for
 the data from Honda et al., Westin et al., Sneden et al., Hill et al.,
 Johnson and Bolte, and Christlieb et al. are
 typically $\pm$0.15, $\pm$0.3, $\pm$0.3, $\pm0.07$, $\pm$0.07 and
 $\pm$0.22, respectively. }
\makebox[4.5cm]{\def\arraystretch{0.3}
\begin{tabular}[t]{ccccccc}\hline
{object}&{$\log_{\epsilon}$(Eu)}&{$\log_{\epsilon}$(Th)}&{$\log_{\epsilon}$(U)}&{Th/Eu}&{U/Eu}&{reference}\\ \hline
{HD6268}&{-1.56}&{-1.93}&{}&{0.69}&{}&{\citet{honda04}}\\ 
{HD110184}&{-1.91}&{-2.50}&{}&{0.55}&{}&{\citet{honda04}}\\ 
{HD115444}&{-1.97}&{-1.97}&{}&{1.00}&{}&{\citet{honda04}}\\ 
{HD186478}&{-1.34}&{-1.85}&{}&{0.60}&{}&{\citet{honda04}}\\ 
{CS30306-132}&{-1.02}&{-1.12}&{}&{0.90}&{}&{\citet{honda04}}\\ 
{CS22892-052}&{-0.86}&{-1.42}&{}&{0.57}&{}&{\citet{honda04}}\\ 
{CS31082-001}&{-0.59}&{-0.92}&{}&{0.72}&{}&{\citet{honda04}}\\ 
{HD115444}&{-1.63}&{-2.23}&{}&{0.55}&{}&{\citet{westin00}}\\ 
{CS22892-052}&{-0.95}&{-1.57}&{}&{0.54}&{}&{\citet{sneden03}}\\ 
{CS31082-001}&{-0.76}&{-0.98}&{-1.92}&{0.80}&{0.31}&{\citet{hill02}}\\ 
{HD108577}&{-1.48}&{-1.99}&{}&{0.60}&{}&{\citet{johnson01}}\\ 
{HD186478}&{-1.56}&{-2.26}&{}&{0.50}&{}&{\citet{johnson01}}\\ 
{HD115444}&{-1.81}&{-2.36}&{}&{0.58}&{}&{\citet{johnson01}}\\ 
{BD+8 2856}&{-1.16}&{-1.66}&{}&{0.61}&{}&{\citet{johnson01}}\\ 
{M92 VII-18}&{-1.48}&{-1.95}&{}&{0.63}&{}&{\citet{johnson01}}\\ 
{M15 K341}&{-0.88}&{-1.47}&{}&{0.55}&{}&{\citet{sneden00}}\\ 
{M15 K462}&{-0.61}&{-1.26}&{}&{0.52}&{}&{\citet{sneden00}}\\ 
{M15 K583}&{-1.22}&{-1.70}&{}&{0.62}&{}&{\citet{sneden00}}\\ 
{CS29497-004}&{-0.45}&{-0.96}&{}&{0.60}&{}&{\citet{christ04}}\\ \hline
\end{tabular}
}
\label{table4}
\end{table}

\acknowledgments
This work has been supported in part by Grants-in-Aid for Scientific
Research (13640313, 17540275) and is for Specially Promoted Research
(13002001) of the Ministry of Education, Science, Sports and Culture of
Japan, and The Mitsubishi Foundation. This work has also been supported 
in part by the U.S. National Science Foundation Grant No.\ PHY-0244384
and by the University of Wisconsin Research Committee with funds granted
by the Wisconsin Alumni Research Foundation. A.B.B. gratefully
acknowledges the 21st Century for Center of Excellence Program
``Exploring New Science by Bridging Particle-Matter Hierarchy'' at Tohoku
University for financial support and thanks the Nuclear Theory Group at
Tohoku University for their hospitality.

\begin{figure}
\centering
  \rotatebox{0}{\includegraphics[width=13.5cm,height=8.5cm]{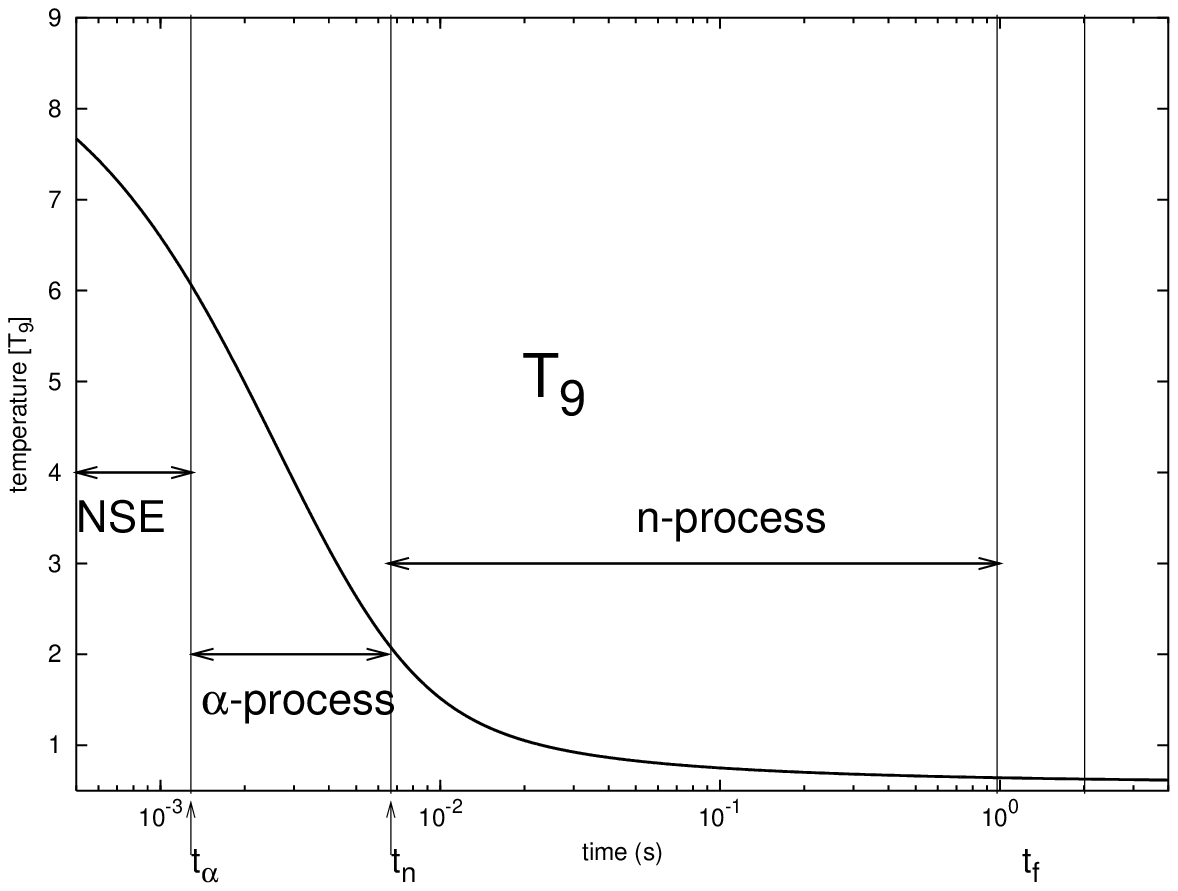}}
  \rotatebox{0}{\includegraphics[width=13.5cm,height=8.5cm]{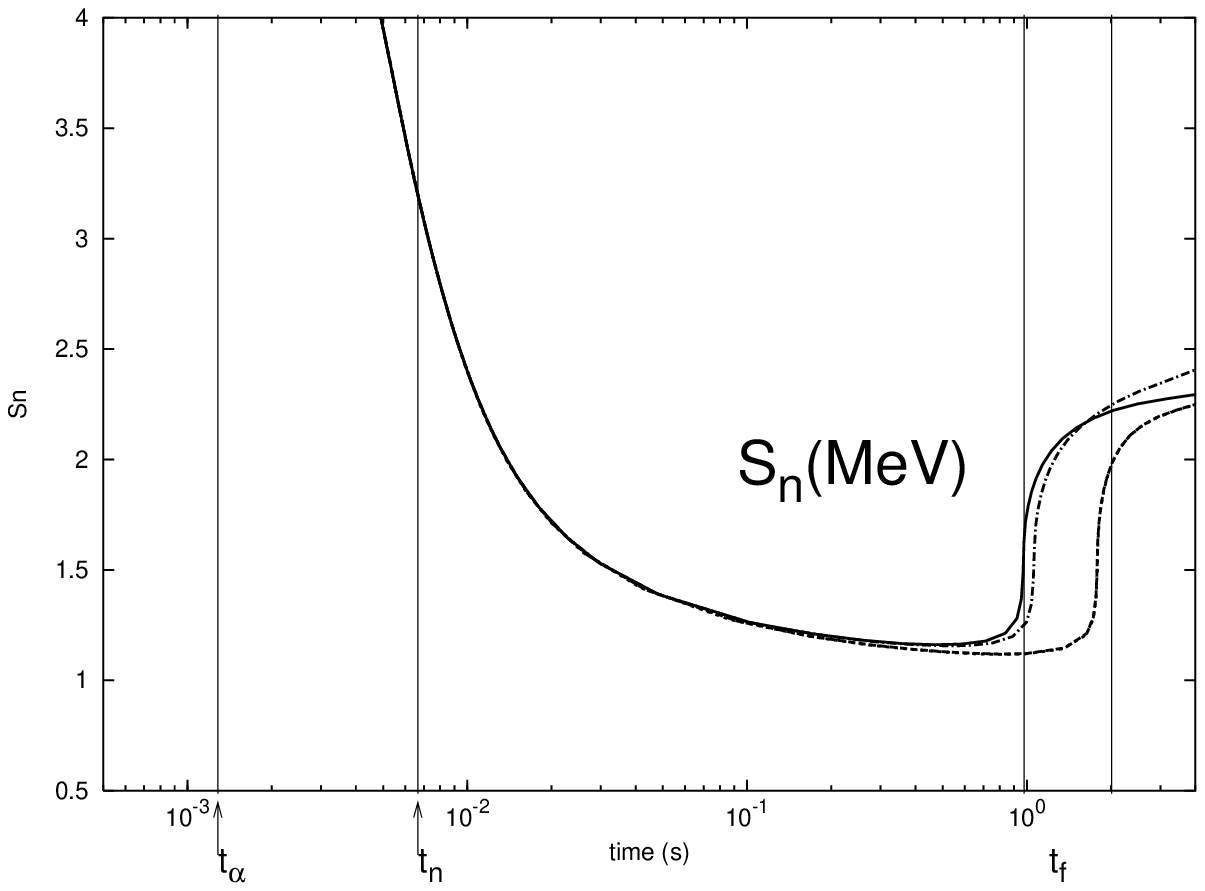}}
  \rotatebox{0}{\includegraphics[width=13.5cm,height=8.5cm]{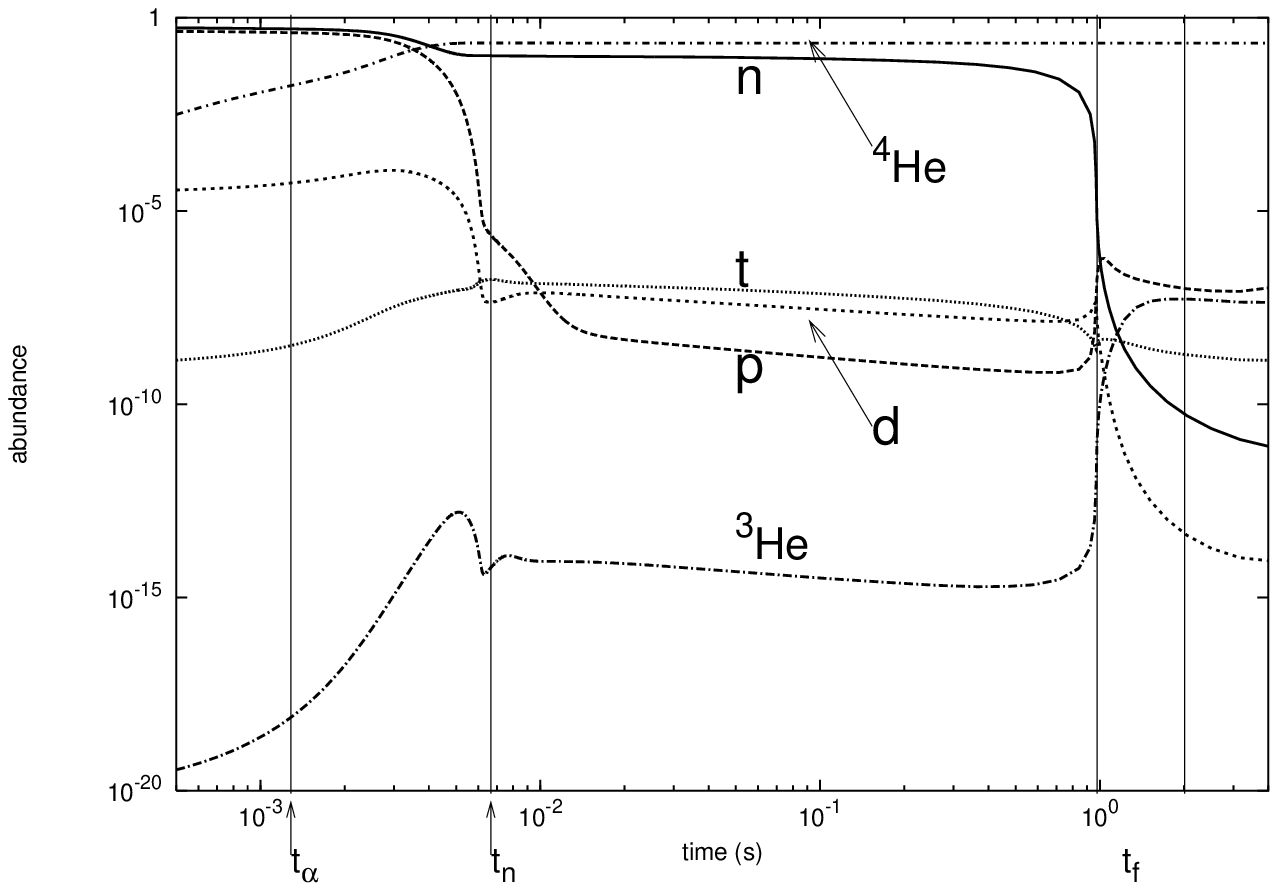}}
\end{figure}\clearpage

\begin{figure}
\centering
\caption{Time evolution of temperature $T_{9}$ (top panel),
  neutron-separation energy $S_{n}$ [MeV] (middle panel),
  and abundance of several light elements (bottom panel)
  in Otsuki flow model \citep{otsuki00}.
  Temperature and light element abundances are for
  no neutrino cutoff, and neutron-separation energies are for four
  different neutrino cutoff times 0.001s (dashed line),
0.005 s (dotted line),
0.1 s (dash-dotted line),
and $\infty$ (solid line).
  Nucleosynthesis starts from NSE, which is taken over by the $\alpha$-process
  during $t_{\alpha} \leq t \leq t_n$, and finally
  the neutron-capture process proceeds during $t_n \leq t \leq t_f$
  by the time of  freezeout of r-process
  when $S_{n}$ drastically increases at $t_f \sim 1-2$ s.
See text for more details. }
\label{fgcut-temp}
\end{figure}\clearpage

\clearpage

\clearpage

\clearpage

\clearpage

\end{document}